\documentclass[conference]{IEEEtran}

\usepackage{times}
\usepackage{graphicx}
\usepackage{multirow}
\usepackage{subfigure}
\usepackage{url}
\usepackage{amssymb,amsmath}
\usepackage{algorithm}
\usepackage{algpseudocode}
\usepackage{dblfloatfix}
\usepackage{todonotes}

\IEEEoverridecommandlockouts

\newif\ifpaper

\paperfalse

\ifpaper
\title{OpenGeoBase: Information Centric Networking meets Spatial Database applications}
\else
\title{OpenGeoBase: Information Centric Networking meets Spatial Database applications - Extended Version}
\fi

\author{\IEEEauthorblockN{Andrea Detti,  Nicola Blefari Melazzi, Michele Orru, Riccardo Paolillo, Giulio Rossi}
	\IEEEauthorblockA{Electronic Engineering Dept., University of Rome "Tor Vergata", Italy\\ Consorzio Nazionale Interuniversitario per le Telecomunicazioni (CNIT) \\ 
		Email: name.surname@uniroma2.it}
	\vspace{-10pt}
	\thanks{This research was partly funded by the EU H2020 Bonvoyage and EU-JP H2020 ICN2020 projects}
	}

\begin{document}
\maketitle

\begin{abstract}
This paper explores methodologies, advantages and challenges related to the use of Information Centric Networking (ICN) for realizing distributed spatial databases. Our findings show that the ICN functionality perfectly fits database requirements: routing-by-name can be used to dispatch queries and insertions, in-network caching to accelerate queries, and data-centric security to implement secure multi-tenancy.
We present an ICN-based distributed spatial database, named OpenGeoBase, and describe its design choices. Thanks to ICN, OpenGeoBase can quickly and efficiently provide information to database users; easily operate in a distributed way, using many database engines in parallel; secure every piece of content; slice resources, so that several tenants and users can concurrently and independently use the database.
We also show how OpenGeoBase can support a real world Intelligent Transport System application, by enabling discovery of geo-referenced public transportation information.

\end{abstract}

%\begin{IEEEkeywords}
%Information Centric Networks,Spatial Database
%\end{IEEEkeywords}

%\category{C.2.2} {COMPUTER-COMMUNICATION NETWORKS} {Network Protocols}
%\category {C.4} {Performance of Systems} {Modeling techniques}
%\terms{Algorithms, Performance, Theory}
%\keywords{Information Centric Networks, Spatial Database, Applications, Implementations}

\section{Introduction}
Spatial databases are storage systems specialized in managing data items related to \emph{space} \cite{SDBintroduction}. The space of interest may be real or virtual geographical spaces and spatial data represent information about the location and shape of contained geometric objects. These objects can be points, multi-points, polygons, lines, etc.. Spatial databases provide spatial query capabilities, e.g. proximity queries that return objects close to a point, polygon queries that return objects within polygons, etc. Spatial databases are used for many applications, including Geographic Information System (GIS), navigation software, journey planners, etc. 

A database system can grow up by adding processing and storing resources, such as memory, storage space or number of CPUs. New resources can be deployed either in existing database servers (vertical scaling), or in new servers (horizontal scaling or sharding), thus realizing a distributed system. The distributed system is made up by a cluster of servers and by an interconnecting network, where a routing function dispatches queries and insertions towards servers \cite{du2007sd}. 

In this paper, we propose the use of an Information Centric Network (ICN) \cite{jacobson2009networking}, namely NDN \cite{ndn}, to realize a distributed spatial database system, named OpenGeoBase (OGB). OpenGeoBase stores geo-referenced information in GeoJSON format, supports range-queries and multi-tenancy. It exploits ICN functionality as follows:
\begin{itemize}
\item routing-by-name to dispatch queries and data insertions, 
\item in-network caching to speed up query responses, avoiding the database engine processing, 
\item data-centric security and trust model \cite{ndntrust} to support secure multi-tenancy.
\end{itemize}

In the next sections we present specific procedures and algorithms for querying and inserting geo-referenced information in the distributed database, highlighting related issues and describing how we exploit ICN's assets, while limiting possible related drawbacks. 
We also show how OpenGeoBase can support a real application \cite{webapp} in the field of Intelligent Transport Systems, to collect, discovery and make available transport information expressed in the format of GTFS transit feed files \cite{GTFS}. Our open-source implementation \cite{ogb-software} is based on the NDN software \cite{ndn}, but CCN \cite{ccnx} could be used as well.

\section{Related works and contributions}
We assume the reader acquainted with the NDN/CCN architectures  \cite{jacobson2009networking} and we mainly discuss related works regarding spatial databases and ICN application for spatial services.

Spatial databases are usually implemented as extension/plug-in of a general purpose database management system (DBMS). They may be based either on a relational or on a NoSQL DBMS model. PostGIS is the spatial extension of the popular open-source PostgreSQL relational DBMS, which adds support for geographic objects, allowing spatial SQL queries using several geometries. Being a SQL database it is hard to distribute PostGIS on different servers and performance problems show up when the database size increases. Other (not open-source) SQL databases with spatial extension are Oracle SQL and Microsoft SQL.

For large data-set (Big Data) NoSQL databases are more and more replacing relational ones, since they can be easily distributed over different servers. Distribution is carried out by grouping the data by a "sharding-key" and using such key to partition the data set among the servers. MongoDB \cite{mongo}, BigTable (by Google), Cassandra (by Facebook), CouchDB (by Apache) are a popular NoSQL databases with spatial support. With respect to a relational DBMSs, a NoSQL database has fewer instruments for managing query geometry. However, for applications  requiring many simple read/write operations on huge data sets, NoSQL databases  are deemed to perform better than relational ones since can be easily distributed. 

Literature on the use of ICN for spatial services is very limited, does not concern database applications, and is mainly focused on vehicular services. In \cite{grassi2015navigo} authors propose to label geographical areas with names and use a routing-by-name schemes to carry out location based forwarding in VANET environment. We label geographical areas too, but with the different goal of routing spatial queries. In \cite{drira2014ndn} authors propose to use ICN to dispatch queries (Interest messages) towards nodes of a V2X network. The Interests contain conditions in the name; many nodes (vehicles, road side units, etc.) may receive an Interest message (flooding) but only nodes that have data satisfying the conditions will send back an answer. We use ICN for query dispatching too, but do not embed conditions inside an Interest to increase cache hit probability and speedup database server processing; we route Interest only towards the database server that can actually serve the query (no flooding), avoiding useless query processing on other databases, thus limiting system load and latency.

%\section{Advance with respect state of art}
In this framework the main contributions of this paper is a first exploitation of ICN in the framework of distributed spatial databases, while identifying specific issues and showing practical, and implemented solutions. Specific contributions in terms of functionality and performance are the following ones.  

In terms of functionality, we classify OpenGeoBase as a NoSQL database since it does not follow a relational model but well support horizontal scaling.  With respect to the other NoSQL spatial databases previously mentioned, OGB data distribution approach is based on a geographic partitioning of the data set among different servers, while routing queries only towards those that can actually contribute to the answer \cite{laurini1998spatial}, avoiding soliciting useless servers. And this is possible exploiting the ICN routing-by-name. For instance, with a proper configuration of the ICN routing plane OGB can be used to realize an international federated database system, in which different servers manage information of different nations. 
%As use case, European Parliament Directive 2010/40/EU for the deployment of Intelligent Transport Systems requires the Member States to set up a national access point for accessing to road and traffic data. Such national access point may actually be a database server of a OGB federated system. 
Another benefit of OpenGeoBase with respect to the other NoSQL databases regards multi-tenancy and data validation functionality, which are offered off-the-shelf, thanks to ICN data-centric security and trust chain model. OpenGeoBase workflow has two distinct actors: users and tenants. Every user of a tenant can write its own data, read data of other users of the same tenant, being sure of data provenance and integrity.      

In terms of performance, even though OpenGeoBase has an initial proof-of-concept implementation, we measured values of range-query latency in the order of hundreds of milliseconds that in some cases are lower than MongoDB ones, and this is a promising starting point for further performance optimization.
\section{OpenGeoBase}
In this section we present the services offered by OpenGeoBase and how these services are realized exploiting NDN functionality. Our design choice comes out from a real implementation available in \cite{ogb-software}. 
\label{s:opengeobase}

\subsection{Offered services}

OpenGeoBase (OGB) is a multi-tenant spatial database. A tenant is a principal that rents a slice of OGB storing space and makes it available to its users. Users can stores geo-referenced information, structured as GeoJSON Feature objects \cite{geojson}. For instance,  a geo-referenced shopping application could use the following GeoJSON format to represent the presence of a Starbucks shop in GPS coordinate 12.51133E, 41.8919N.  

\vspace{3pt}
\noindent\texttt{\small
	\{"type": "Feature", 
	"geometry": \{"type": "Point","coordinates": [12.51133, 41.8919]\},
	"properties": \{"oid" : 1234, "tid" : "Foo", "uid": "Alice", "cid": "ShopApp", "shop-name": "Starbucks", "shop-type" : "coffeehouse" \}\}\}	
	}
\vspace{3pt} 
\noindent  The oid, tid, uid and cid properties are mandatory. The oid is a unique identifier of the GeoJSON object (a random nonce), tid and uid are the identifiers of the tenant and of the user. Tenant data are grouped in \emph{collections} and cid is the collection identifier. The shop-name and the shop-type (and other possible ones) are customizable application properties.     

Users can carry out either inclusion or intersect range-queries for obtaining all the GeoJSON objects which are completely (inclusion) or partially (intersect) contained in the range-query area, i.e. a 2D bounding box. 

For scaling purposes, the system administrator(s) can deploy a distributed set of database servers (or engines), each one dedicated to store data related to a zone of the world.

\subsection*{Spatial Indexing}
OpenGeoBase uses a three-levels spatial indexing grid, aligned with world parallels and meridians (fig. \ref{f:tilehier}). Grid regions are called \emph{tiles} and we motivate the choice of such indexing schemes in appendix I \ifpaper of \cite{ext} \fi. A level-0 tile of the grid contains all world points having the same longitude and latitude values up to the dot, e.g. the level-0 tile (12,41) contains all points whose longitude and latitude start with 12 and 41, respectively. A level-$n$ tile of the grid, where $n=1,2$, contains all points having the same latitude and longitude value up to the $n$th decimals, e.g. the level-2 tile (12.51,41.89) contains all points whose latitude and longitude start with 12.51 and 41.89, respectively. In doing so the \emph{level-ratio} of the grid is 100, i.e. each tile of level-$n$ is formed by 100 tiles of level-$(n+1)$. For latitudes close to the equator, the area of a level-0 tile approximates a square of 100x100 km, a level-1 tile a square of 10x10 km, and a level-2 tile a square of 1x1 km. A level-$n$ tile is identified by an ICN name-prefix, called \emph{tile-prefix}, whose structure is:

\vspace{3pt}
\noindent\texttt{\small ndn:/OGB/lng(0)/lat(0)/lng(1)lat(1)/...\\
	/lng(n)lat(n)/GPS-ID}
\vspace{3pt}

\noindent where, lng(0) and lng(x) are the integer and the $x$th decimal value of the GPS longitude of the tile. The same relation holds for latitude values, e.g., the level-2 tile (12.51,41.89) has the name prefix ndn:/OGB/12/41/58/19/GPS-ID.

\begin{figure}[t]
	\centering
	\includegraphics[scale=0.18]{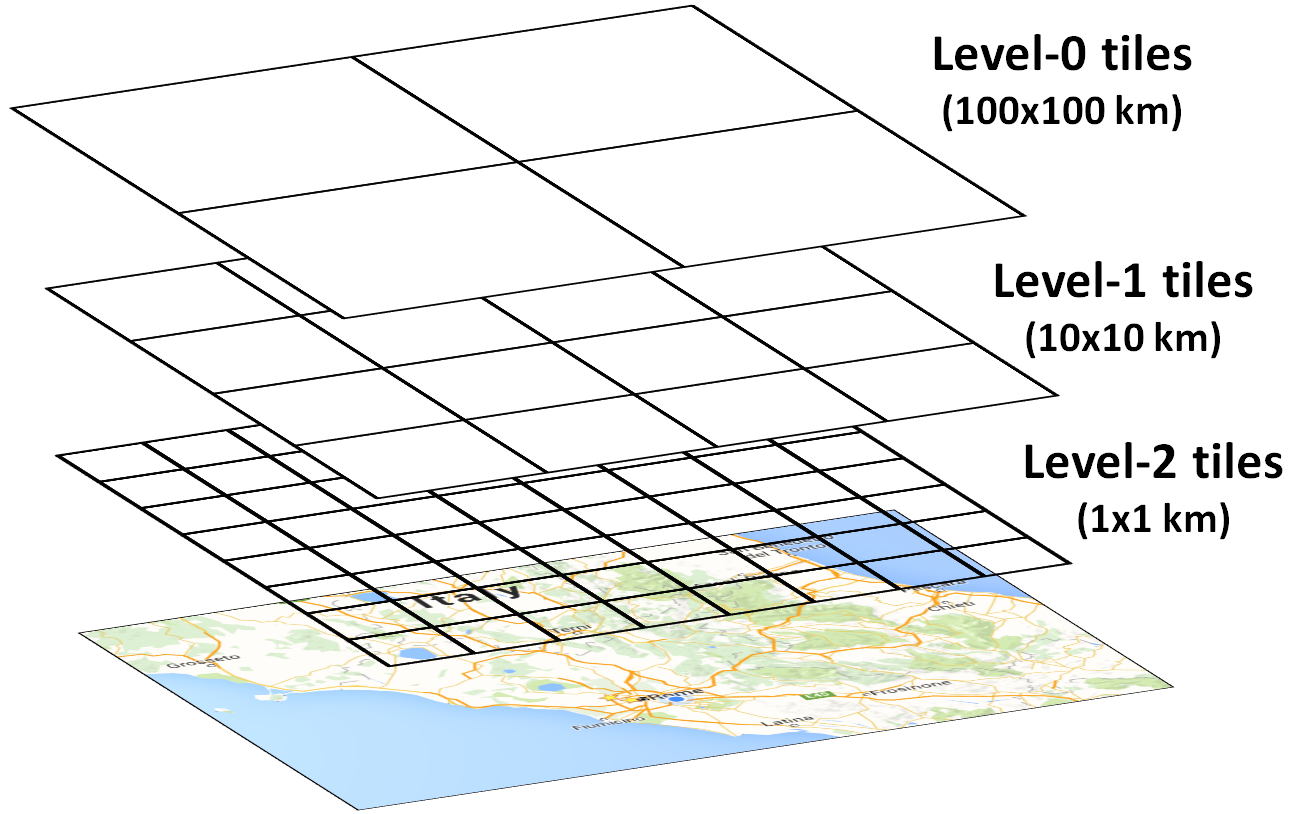}
	\caption{Tile hierarchy example for level-ratio 4 }
	\label{f:tilehier}
	\vspace{-12pt}
\end{figure}

\subsection{Internal data structures}
\begin{figure}[t]
	\centering
	\includegraphics[scale=0.28]{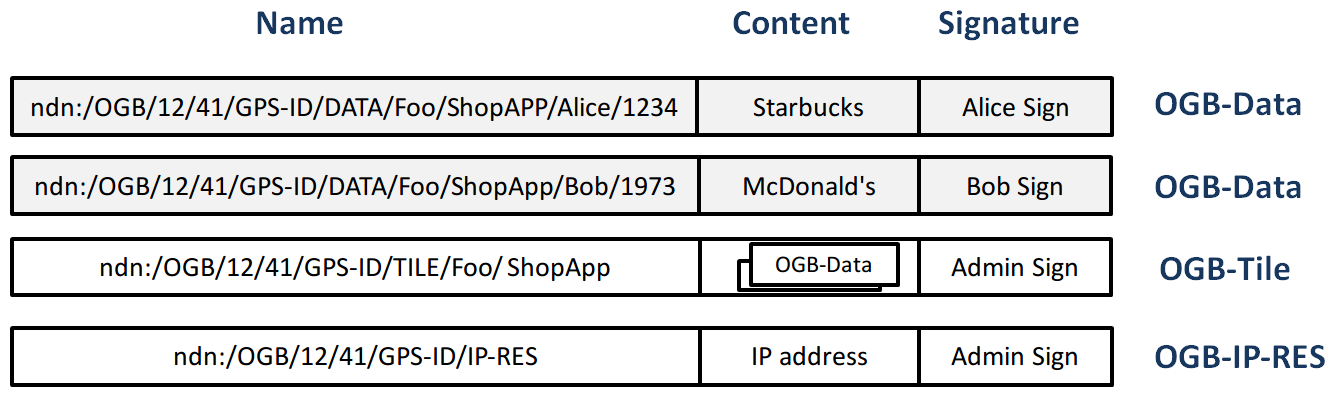}
	\caption{OGB-Data , OGB-Tile and OGB-IP-Res structures (no segmentation) }
	\label{f:datastructure}
	\vspace{-12pt}
\end{figure}

OpenGeoBase uses three kinds of ICN data structures for internal purposes, namely: \emph{OGB-Data}, \emph{OGB-Tile} and \emph{OGB-IP-Res}, reported in fig. \ref{f:datastructure}.

An OGB-Data is used to index and link/store a GeoJSON object as follows. A GeoJSON object intersects different tiles of the grid, at any level. For each intersected tile, there exist in the system an OGB-Data item, which contains either the GeoJSON object or a reference to another ICN content containing the actual GeoJSON object. In doing so, a GeoJSON object is stored only one time in the system. An OGB-Data is actually an ICN content, in case segmented in a sequence of ContetObjects, and identified by a unique name whose scheme is the following:  

\vspace{3pt}
\noindent\texttt{\small ndn:/tile-prefix/DATA/tid/cid/uid/oid}
\vspace{3pt}    

For instance, the GeoJSON of Starbucks aforementioned intersects the level-2 tile $(12.51,41.89)$. Thus, there exist in the system an OGB-Data content whose name is ndn:/OGB/12/41/58/19/GPS-ID/DATA/Foo/ShopApp/Alice/1234 and whose payload is the GeoJSON object. We observe that the same GeoJSON object also intersect the $(12.5,41.8)$ and the $(12,41)$ level-1 and level-0 tiles, respectively. Thus, other two OGB-Data contents exist whose content is a reference to ndn:/OGB/12/41/58/19/GPS-ID/DATA/Foo/ShopApp/Alice/1234.       

An OGB-Tile is an OGB-Data container used to support a \emph{tile-query}, i.e. a query requesting all the GeoJSON objects of a collection which intersects a given tile. OGB-Tile is actually an ICN content, in case segmented, with a unique name and whose payload is the set of OGB-Data items related to the queried tile. A tile-query can be carried out through a standard ICN GET primitive (e.g. ndngetfile) based on Interest/ContentObjects message exchange. The OGB-Tile naming scheme is:

\vspace{3pt}
\noindent\texttt{\small ndn:/tile-prefix/TILE/tid/cid} 
\vspace{3pt}

OGB-Tile contents are cached by ICN nodes and popular tile-queries can be quickly satisfied from caches, without involving database processing. To avoid the presence of stale data in the cache due to data updates, the FreshnessPeriod of the OGB-Tile ContentObjects can be properly configured by the tenant. Otherwise, if a more reactive cache update approach is needed, other network mechanisms can be introduced, even though such important research issue, namely ICN caching for database application, is left for future works.  

The OGB-IP-Res is an ICN content used by the Data Insert procedure to derive the IP address of the database engine handling a given tile-prefix. The related naming scheme is:

\vspace{3pt}
\noindent\texttt{\small ndn:/tile-prefix/IP-RES}   
\vspace{3pt}

\subsection {System Architecture} 

\begin{figure}[t]
	\centering
	\includegraphics[scale=0.2]{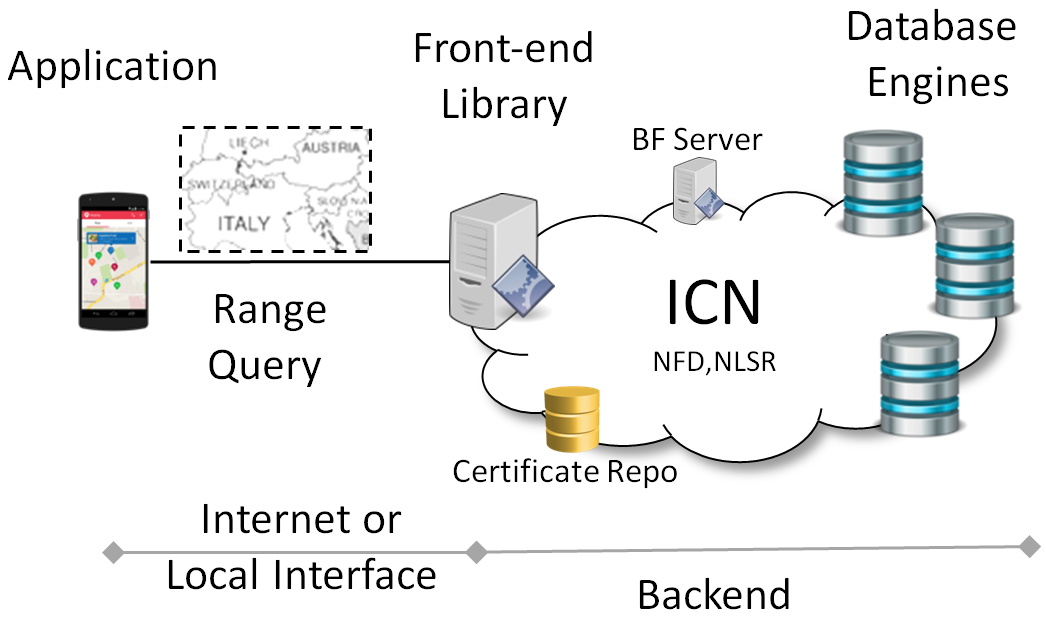}
	\caption{OpenGeoBase Architecture}
	\label{f:arch}
	\vspace{-12pt}
\end{figure}

\begin{figure}[t]
	\centering
	\includegraphics[scale=0.22]{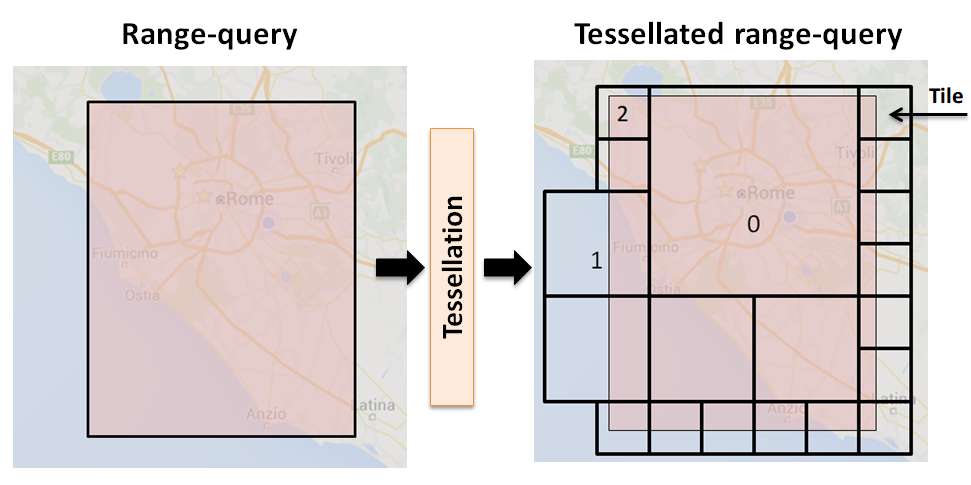}
	\caption{Constrained tessellation with $k=19$}
	\label{f:tesseling-concept}
	\vspace{-12pt}
\end{figure}

Fig. \ref{f:arch} reports the OpenGeoBase distributed architecture whose components are described in next subsections.

\subsubsection{Database engines} a database engine is an ICN repository that serves a subset of tiles, by storing the related OGB-Data contents and processing queries of OGB-Tile. We implemented the database engine as an extension of NDN repo-ng software. The extension is mainly used for the on-demand building of OGB-Tile contents, which requires to select and package in the OGB-Tile payload all the stored OGB-Data whose tile-prefix is equal to the one of the requested OGB-Tile. To speed-up this search we inserted other specific tables in the SQLite DBMS of repo-ng, as better described in the appendix II \ifpaper of \cite{ext} \fi.   

\subsubsection{Front-end library and BF server} the front-end library handles range-queries and insertion of GeoJSON objects coming from an application, by using the procedures described in the next section. A Bloom Filter (BF) server can be used to speed up queries. The application and the front-end library can run in the same device (fat-client approach) or in different devices connected by the Internet, e.g. the front-end can run within an HTTP application-server. To avoid bottlenecks, many application-servers can be deployed and handled by a load balancing function, such as DNS or reverse-proxy.       

\subsubsection{NFD, NLSR and Certificate Repo} each component of the OGB backend (fig. \ref{f:arch}) uses NDN Forwarding Daemon (NFD) for routing-by-name tile-queries (OGB-Tile Interest), caching tile-query responses (OGB-Tile ContetObjects), secure tile-queries and data insertions (OGB-Data ContetObjects). Moreover, each component uses NLSR routing protocol \cite{hoque2013nisr} to configure the NFD Forwarding Information Base (FIB). Using NLSR each database engine announces on the routing plane the tile-prefixes of its tiles, thus the ICN is able to route tile-queries towards the proper engine. For security procedures tenant and user NDN certificates are stored in an internal ICN repository.

\begin{figure}[t]
	\centering
	\includegraphics[scale=0.2]{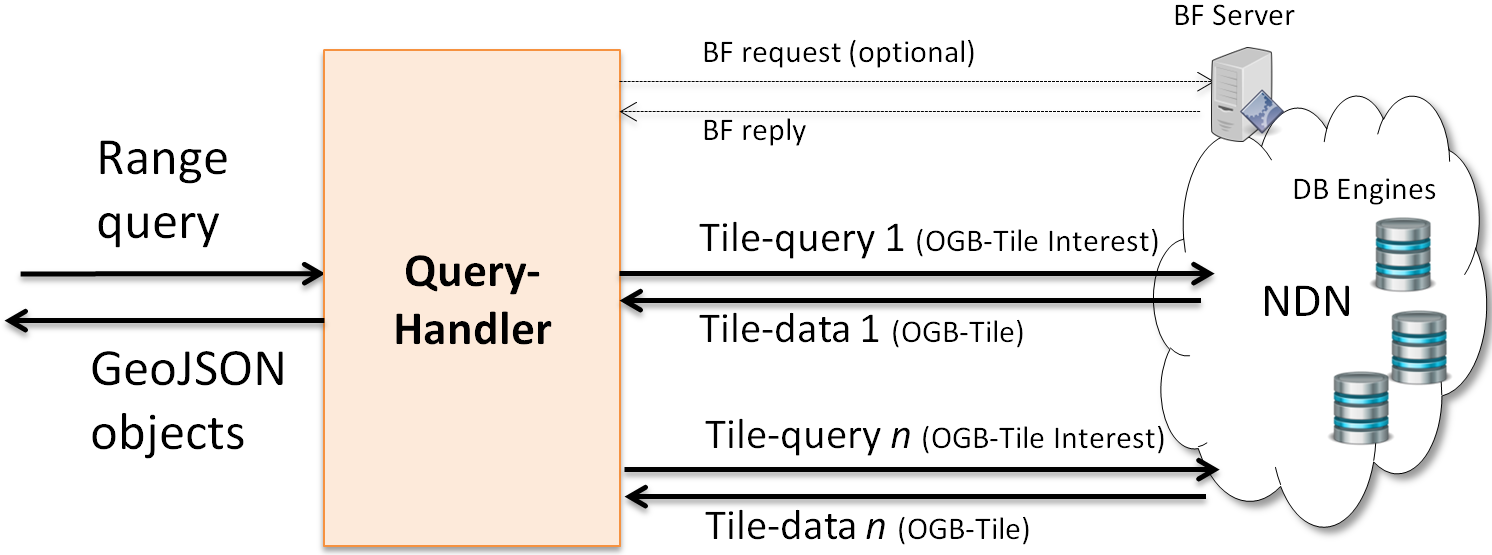}
	\caption{Range-query}
	\label{f:queryhandler}
	\vspace{-12pt}
\end{figure}

\subsection{Procedures}
	
\subsubsection{Range-query} 
\label{s:rq}	
an intersect or include range-query is handled by a query-handler function within the front-end library (fig. \ref{f:queryhandler}). The query-handler resolves the range-query in two phases: tile-querying and post-filtering.

In the tile-querying phase the query-handler covers the range-query area $A$ with set of tiles identified by a \emph{constrained tessellation algorithm}. Then for each tile it carries out a tile-query, i.e. an ICN GET of an OGB-Tile. The set of tiles has a constrained size $k$ and covers an area $B$, which contains the range-query area $A$, but also some extra border space, due the fact that the range-query area $A$ may be not aligned with the grid. For instance, in fig. \ref{f:tesseling-concept} a range-query area is tessellated with $k=19$ tiles of level 0 ,1 and 2. In appendix III \ifpaper of \cite{ext} \fi we developed an heuristic tessellation algorithm that uses as much possible the largest tiles to cover the area $A$, while satisfying the constraint $k$ and limiting the extra border space. When the area $A$ is so large as it is not possible to cover it not even with $k$ level-0 tiles, the constraint can not be respected and the algorithm merely covers the area with the minimum number of level-0 tiles. Reducing the constraint $k$, on the one hand, accelerates the range-query processing time, since less tile-queries are required to satisfy a range-query; on the other hand, it could result in an increase of the extra border area, thus the query-handler is fetching more data than requested one, and this increases the transmission and post-filtering time. Clearly a trade-off is needed, which depends on the volume of geo-referenced information.    

When geo-referenced information are rather sparse, many of the tessellated tiles are void. To avoid time consuming ICN void tile-queries, the query-handler may use a Bloom Filter Service. The Bloom Filter (BF) is loaded with tile-prefixes of not-void tiles. After tessellation, the query-handler tests the BF membership of tessellated tiles and obtains a reduced tessellation set comprising only not-void tiles. OpenGeoBase contains one or more BF servers that handle membership requests, as show in fig. \ref{f:queryhandler}. The price to pay is an additional RTT in the range-query time. Such a cost is worth paying only when geo-referenced data are rather sparse and the use of BF is left as an option. To update the BF, OGB uses a ICN topic-based publish-subscribe approach similar to the one presented in \cite{detti2015exploiting}, but other solutions are feasible too \cite{chen2011copss}. Each database engine has a local Counting Bloom Filter (CBF) with the same buckets of the global BF running on the BF servers. At each data insert/removal the local CBF is updated. When a CBF bucket becomes i) greater than zero or 2) equal to zero, a publication is sent out on a system topic, since the related global BF bucket value could need a 0-1 or 1-0 switch, respectively. The BF servers are subscribers of the system topic, and received publications are combined (OR) to configure bucket values.  

When all tile-query responses are received the second post-filtering phase starts, during which the query-handler unpacks the OGB-Data within the received OGB-Tile contents, verify their validity, extracts the enclosed GeoJSON objects and applies a post-filtering aimed to select those GeoJSON objects that actually intersect with or are included in the range-query area. Post-filtering is necessary since the tile-querying phase may return more objects than requested ones, e.g. due to the tessellation extra border space.

\subsubsection{Data Insertion}
\label{s:data-insert}
\begin{figure}[t]
	\centering
	\includegraphics[scale=0.2]{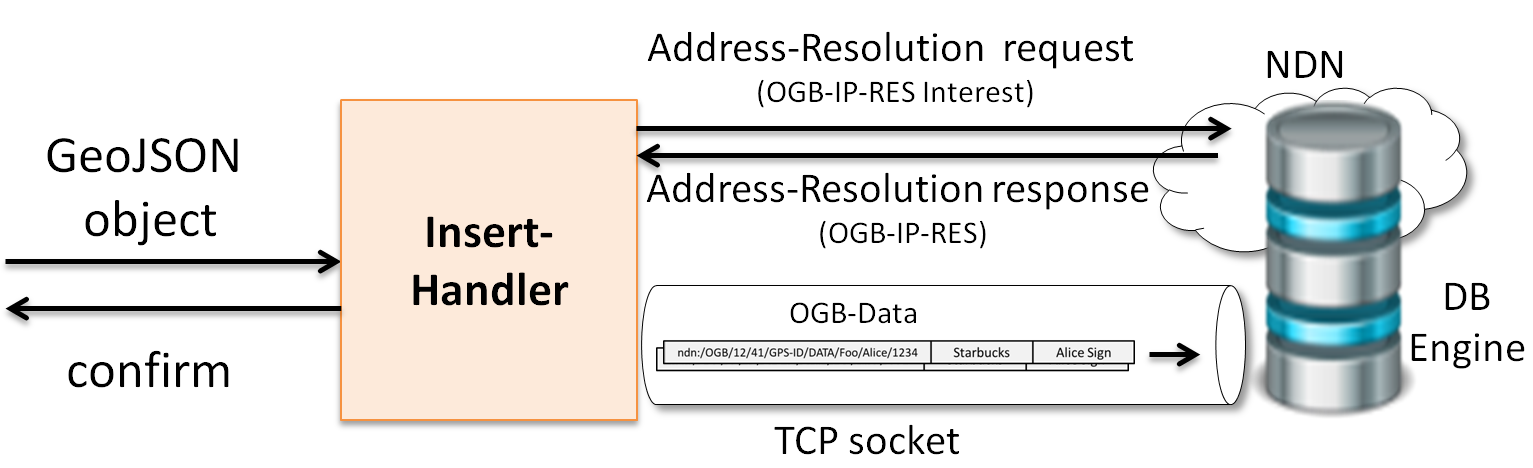}
	\caption{Data Insertion}
	\label{f:datainsert}
	\vspace{-12pt}
\end{figure}

an insertion of a GeoJSON object is handled by an insert-handler within the front-end library (fig. \ref{f:datainsert}). The insert-handler parses the GeoJSON object, creates the related OGB-Data items for each intersecting tile, and push them into the responsible database engines using a "mixed" TCP/IP-ICN procedure, motivated as follows.

ICN natively provides pull services and literature mainly proposes two approaches for pushing data. The first one is used by the NDN Repo Insertion Protocol \cite{ndn}, however its adoption in our database application would imply to set up on-demand an ICN route towards the end user applications requiring data insertion, giving to these applications the power to modify the ICN routing plane, possibly creating or facilitating security and scalability issues of the back-end. For this reason, we do not adopt this approach. The second kinds of push approaches \cite{franccois2013ccn}\cite{carzaniga2011content} require changing of NDN (CCN) forwarding and the introduction of a new "Push" message routed-by-name as an Interest messages. In this version of OGB we preferred to avoid these ICN architectural changes and do not adopt these approaches. Consequently, while looking forward to having an "official" ICN/NDN/CCN push service implementation not requiring routes towards producers, we temporary resorted to a mixed TCP/IP-ICN push approach, which exploits the TCP-bulk-insert procedure of current NDN repo-ng implementation and our ICN Address-Resolution procedure. By using TCP-bulk-insert, the insert-handler sets up a TCP connection with a database engine and push OGB-Data through it.  However, the insert-handler needs to know the IP address of the database engine that should store the OGB-Data. For such an IP address resolution purpose, we developed an ICN procedure for which the interest-handler sends out an Interest message for a OGB-IP-Res content, whose name contains the tile-prefix of the OGB-Data to be pushed. The ICN network routes-by-name such an Interest towards the responsible database engine, which answers with a content containing its IP address.

\subsubsection{Secure multi-tenancy}
The NDN security library makes it possible to sign Interests and ContentObjects. Signed Interests and ContentObjects contain the name of the certificate to be used to verify the signature (aka Keylocator), so that any device can fetch the certificate, if unavailable locally. An NDN certificate includes the identity name of the owner, its public key, the signature of the certification authority and the related KeyLocator, so realizing a trust chain. When an Interest or a ContentObject is received, it can be validated by using a Validator tool, whose trust schema can be flexibly configured by using rules and trust-anchors \cite{ndntrust}. 

To support multi-tenancy in OpenGeoBase,  exploit the NDN security library as is, thus we do not give many details here, also for lack of space. The trust chain assures that the users of a tenant have a unique identity, released by the tenant; the tenant signs and releases user certificates. The identity of the tenant is provided by the OpenGeoBase administrator that also releases and signs tenant certificates. The public key of the system administrator is the trust-anchor. The system has a dedicated repository of certificates. 

An OGB-Tile Interest message related to a tile of a tenant is signed by the message issuer. By using a proper configuration of the Validator and KeyLocator names, edge NDN nodes (or directly database engines, if no other kind of NDN nodes exist in the back-end, as in our case), accept only OGB-Tile Interest messages whose certificate of the issuer has been released by the tenant of the tile. Consequently, a tile-query regarding data of a tenant is processed only if submitted by a tenant user.

An OGB-Data ContentObject is signed by the data owner. By properly configuring the Validator, the query-handler verifies that a received OGB-Data, related to the owner indicated in the OGB-Data name, is actually properly signed by the owner. Thus, both data integrity and data owner (provenance) are validated. Dually, we configure the Validator of the database engine to insert only OGB-Data ContentObjects whose identity (contained in the OGB-Data name) is the same of the owner of the signing certificate. 

%We observe that using such a configuration to speed-up query and insert security checks the database engines should have the certificates of all the users in their certificate store. Indeed, retrieving remotely at run time the certificate using KeyLocator may dramatically delay query and insert processing.     

\section{Modeling and performance evaluation}
We evaluated some key performance parameters of OpenGeoBase in two environments: a laboratory one, to show general OpenGeoBase performance; an application one, to measure the performance of a real world Intelligent Transport Systems  (ITS) application based on OpenGeoBase. 

\begin{figure*}[ht]
	\centering
	\subfigure[Batch tile-query duration vs. number of tile-queries for 1x1 and 10x10 tiles, in case of 1 and 4 database engines, no caching, no BF]{  
		\includegraphics[scale=0.37]{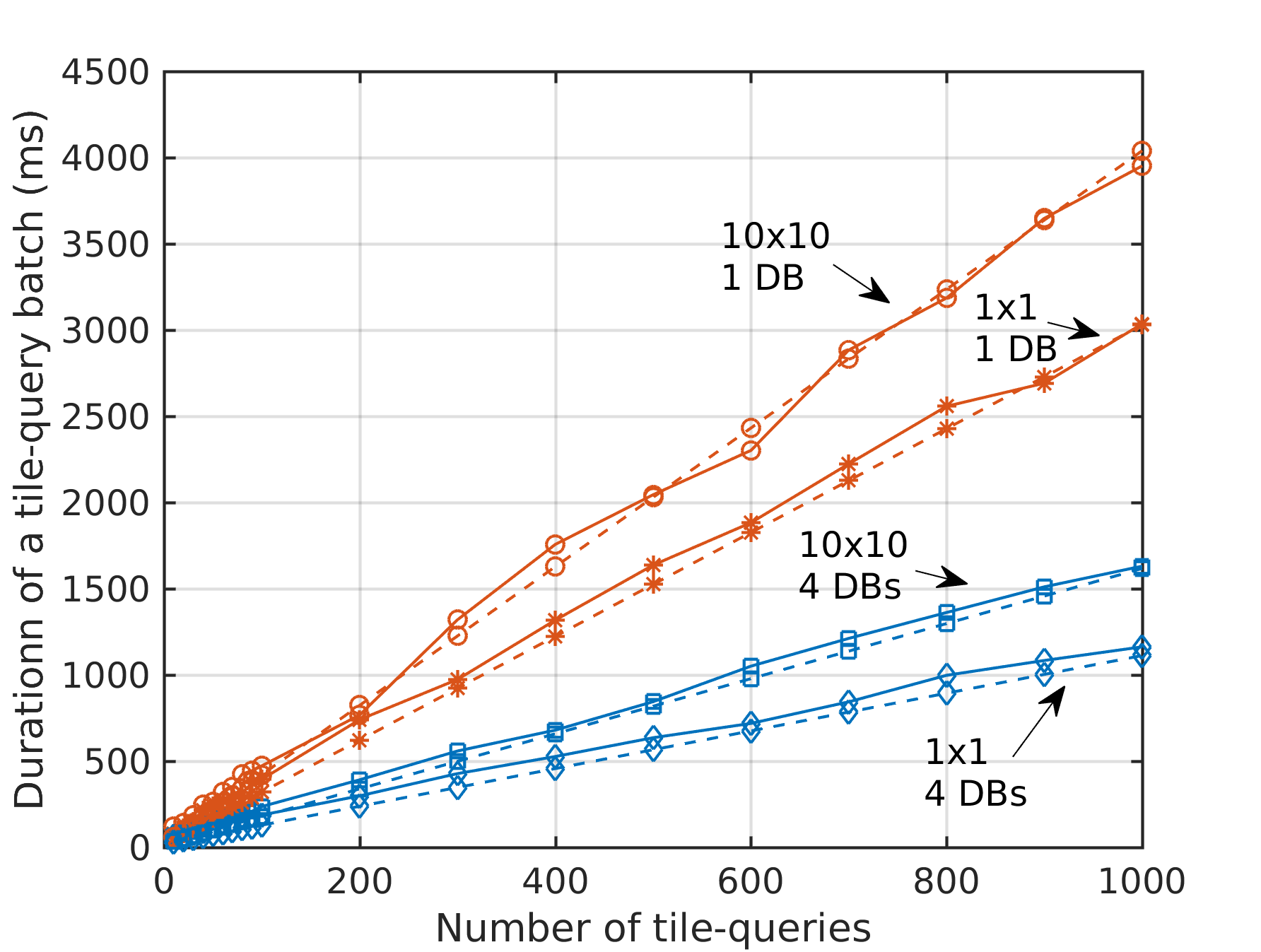}
		\label{f:totalTileQueryTime}
	}
	\subfigure[Duration of a batch of 500 tile-queries time vs. cache hit probability, for 10x10 tiles, in case of 1 and 4 database engines, no BF]{  
		\includegraphics[scale=0.37]{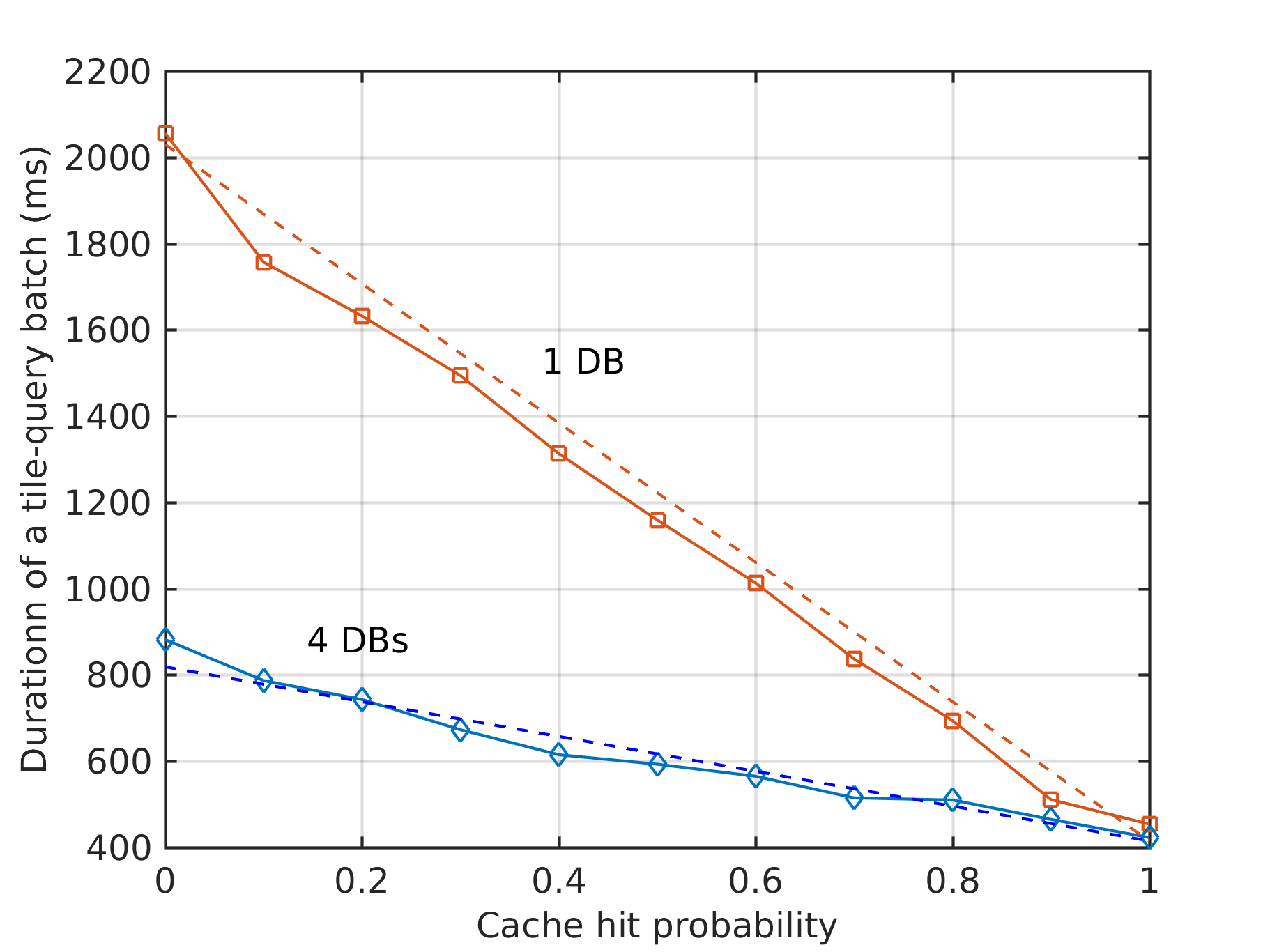}
		\label{f:totalCacheTime}
	}
	\subfigure[Range-query time vs. range query area, 4 database engines, no BF, constrained tessellation with $k$ max-tiles, with and without ICN cache of 5000 items]{  
		\includegraphics[scale=0.37]{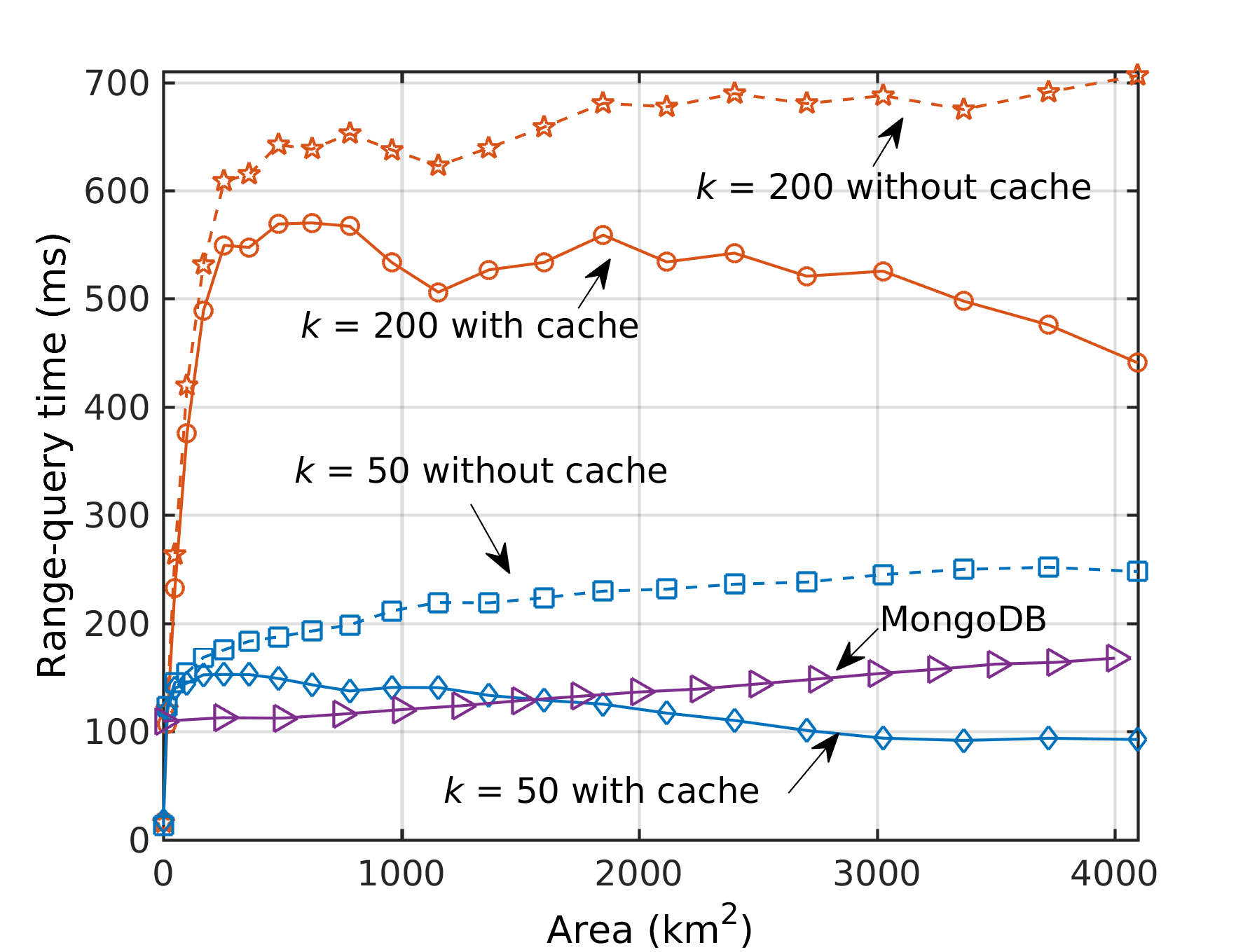}
		\label{f:totalRangeQuery}
	}
	\subfigure[Red dots indicate not-void tiles for the GTFS application]{  
		\includegraphics[scale=0.34]{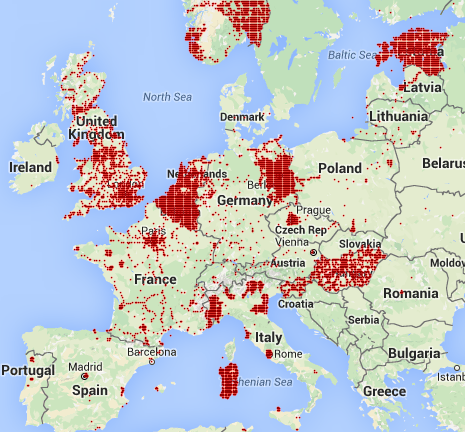}
		\label{f:gtfsword}
	}
	\subfigure[Average duration of range-query, query-tiles batch, tessellation and BF request vs. range-query area for the GTFS application]{  
		\includegraphics[scale=0.38]{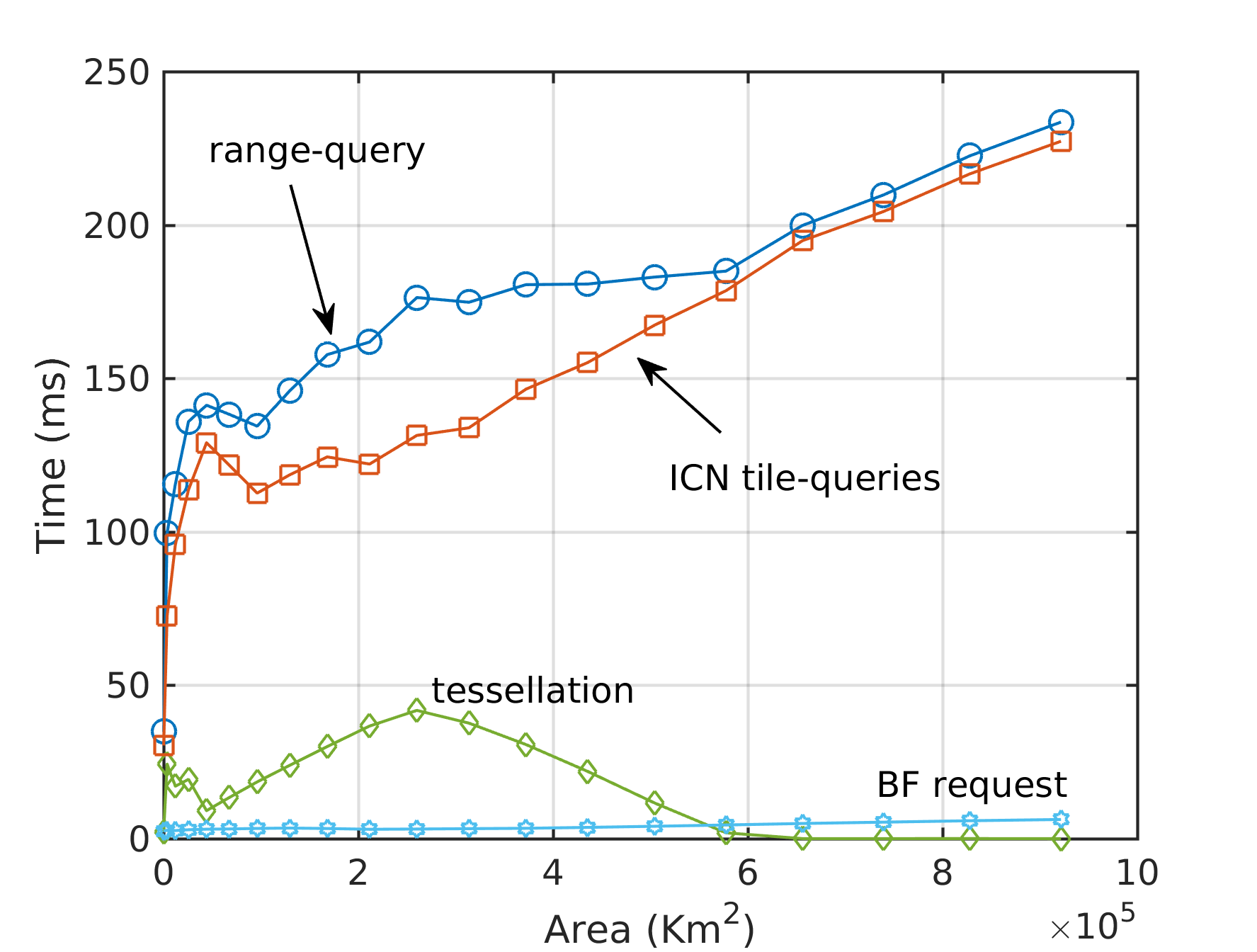}
		\label{f:gtfsqt}
	}
	\subfigure[Average number of tiles vs. range-query area for the GTFS application]{  
		\includegraphics[scale=0.38]{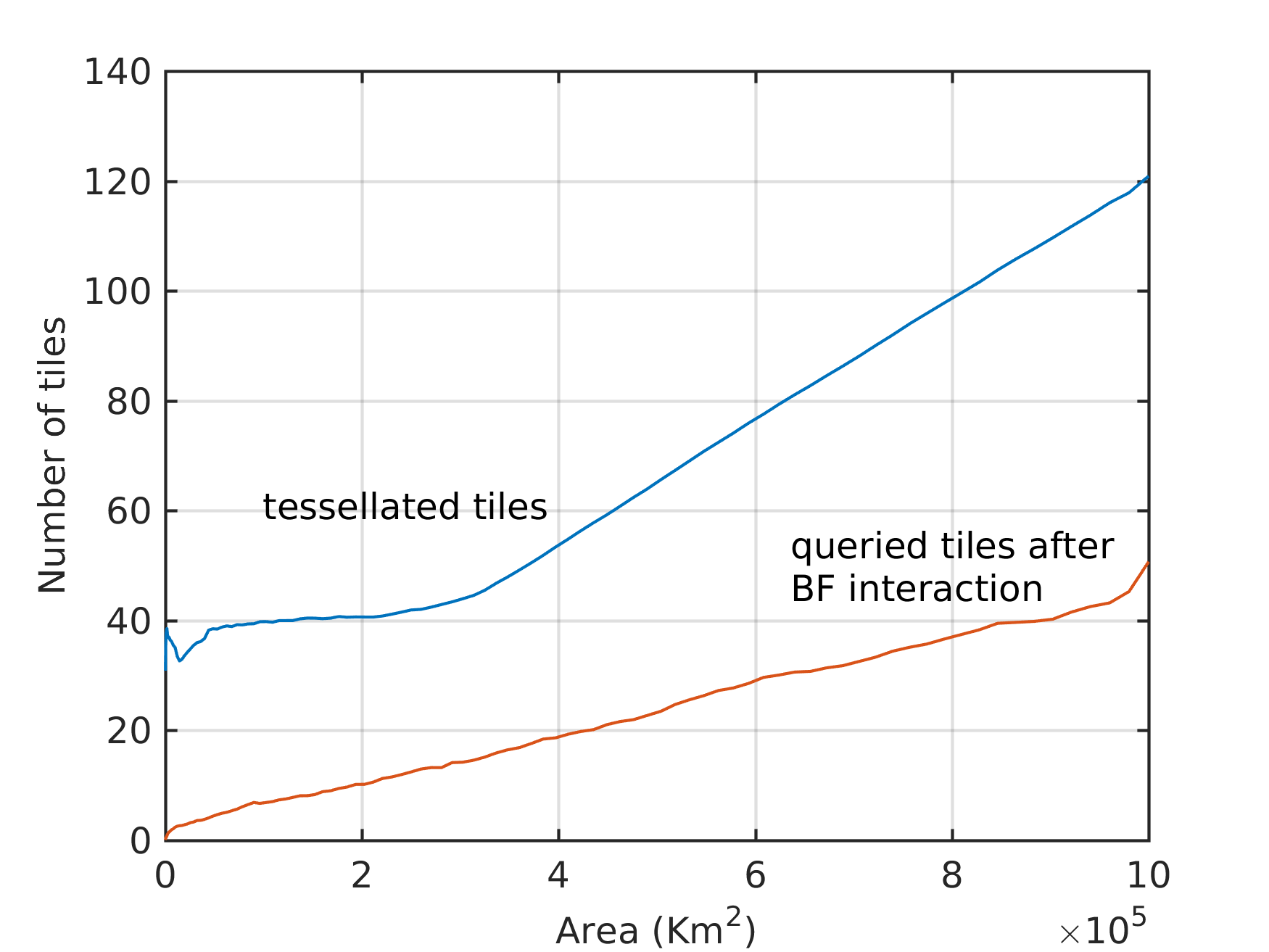}
		\label{f:gtfstiles}
	}
	%\subfigure[Average tessellation stretch for the GTFS application]{  
	%	\includegraphics[scale=0.35]{fig/stretchingRatio_vs_area.png}
	%	\label{f:gtfsstretch}
	%}
	\caption{Performance evaluation results}
	\vspace{-10pt}
\end{figure*}

\subsection{Laboratory tests}
The laboratory environment is formed by: i) a virtual machine (VM) executing the front-end library and a benchmark application producing queries; ii) five VMs running 4 database engines and the Certificate repo. VMs are connected by a Linux bridge with a measured throughput of about 200 Mbit/s at the application layer. 
The database covers an area of 400x400 km, aligned with 100x100 km (level-0) tiles. The data set contain a GeoJSON Point object of 55 bytes for each 1x1 km (level-2) tile. We did not use the Bloom Filter (BF) Server, as spatial data are not sparse. We considered two scenarios: i) one database engine; ii) four database engines, each of them responsible for a different 100x100 km tile. Then we studied the performance of tile-queries and intersect range-queries. 
We also compare performance of intersect range-queries when using MongoDB system whose configuration is formed by 4 database engines (Shards), a query router (Mongos) and a Config Server, running on different VMs.  

Fig \ref{f:totalTileQueryTime} reports the duration of an OGB batch of tile-queries versus the number of queried tiles, without ICN caching, for 1x1 and 10x10 tiles, in case of 1 and 4 database engines. Resulting tile-queries are uniformly distributed on the databases. 
%This figure is interesting since a range-query is actually a batch of tile-queries. 

To better analyze and understand these results we developed a very simple model. The duration of a single tile-query is the sum of processing time ($TQ_p$) and transmission time.  
We experimentally observed that the processing time has a small dependence on the number of returned items ($N_i$) and can be modeled as follows:
\begin{equation}
TQ_p = C_1 + C_2 N_{i}
\label{e:eq1}
\end{equation}
\noindent where $C_1$ is a constant time equal to 3 ms, independent from the number of returned items, and $C_2$ is a constant time equal to 0.008 ms, which multiplies the number of items. Clearly, these numerical values are related to our devices and our data, albeit the formula may hold in general.  
  
The processing of a tile-query takes place in two devices: in the database engine, to extract data from the DBMS and build/sign the OGB-Tile; in the query-handler, to decode and verify the OGB-Tile and the inner OGB-Data items. Consequently, we can approximately decompose the processing time in a fraction $P_{db}$ due to the remote database and a fraction $P_{qh}$ due to the query-handler, where $P_{db}+P_{qh}=1$.     

If we assume to carry out a batch of $N_q$ tile-queries equally split on $N_{db}$ database, the total batch duration $TB$ is equal to:  
\begin{equation}
TB = C_3 + N_q \left((1-H)\frac{P_{db}}{N_{db}}+P_{qh}\right)TQ_p+N_q\frac{N_iD_s}{B_w}
\label{e:eq2}
\end{equation}

\noindent where $C_3$ is a constant value of 20 ms in our configuration, needed by the query-handler for opening and closing the NDN face and for other NDN constant processing; $D_s$ is the size of a single OGB-Data (we are not considering the OGB-Tile overhead); $B_w$ is the throughput at the application layer. The term $P_{db}/N_{db}$ takes into account that we are using $N_{db}$ databases \emph{in parallel}: indeed, horizontally scaling reduces the overall database processing time, but not the query-handler processing. The term $(1-H)$ is the cache miss probability: database processing occurs in case of cache miss. 

The dotted lines reported in fig. \ref{f:totalTileQueryTime} come from eq. \ref{e:eq2} for: $B_w = 200 Mbit/s; P_{qh} = 0.15$; $P_{db}=0.85$; $D_s=55 $ bytes; $H=0$, no cache; $N_i=1$ in case of 1x1 tiles and $N_i=100$ in case of 10x10 tiles. The solid lines come from measurements. For the same kind of tile-queries, e.g. 1x1, increasing the number of databases from 1 to 4 leads to a significant reduction of the duration of the tile-query batch, since we are concurrently using 4 databases. This result is proof that ICN routing-by-name is indeed effective to allow the database to horizontally scale.     
For the same number of databases, the request of 10x10 tiles leads to a greater batch duration, since more data have to be processed and sent back. 
%due to the increase of number of returned items (100 vs 1) and the processing time, as shown in eq. \ref{e:eq1}. 
However, since $C_1>>C_2$, it is much more convenient to use one 10x10 tile rather than ten 1x1 tiles, to query an area of a 10x10 tile. Thus, the choice made for our tessellation approach of using the largest possible tiles is the right one.

Fig. \ref{f:totalCacheTime} shows the impact of ICN in-network caching on the duration of an OGB batch of 500 tile-queries of 10x10 km. Dotted lines come from eq. \ref{e:eq2}. 
Caching has been enabled only on the database engines. By increasing the cache hit probability ($H$), the batch duration decreases to the lowest limit, represented by the sum of the query-handler processing time plus the transmission time. This plot confirms the effectiveness of ICN in-network caching in reducing database query times, since a cache lookup has a negligible duration, with respect to the time needed to access the database engine.  

Fig. \ref{f:totalRangeQuery} shows the (intersect) range-query time versus the range query area, in absence and in presence of ICN caching and for different values of the tessellation constraint $k$. Increasing the area, OGB latency tends to an asymptotic value since the involved number of tile-queries is however limited to $k$. As discussed in section \ref{s:rq}, reducing the max-number of tiles $k$ reduces the range-query time since less tile-queries are necessary. Nevertheless, a greater number of spatial data are filtered out by post-filtering, increasing such a processing time that however remains negligible in our tests. Caching accelerates range-queries, and acceleration improves with the range-query area, since the constrained tessellation tends to use greater tiles. As a result, a smaller universe of ICN contents is circulating in the network, increasing the cache hit probability. 

Fig. \ref{f:totalRangeQuery} also reports MongoDB performance with the  same GeoJSON data set. Considering a reference OGB configuration of $k=50$ with caching, OpenGeoBase latency is greater than MongoDB one for areas lower than 1600 km$^2$. This is mainly due to the different data distribution and query routing approaches. Practically, MongoDB randomly spread spatial objects over the different DB engines. OpenGeoBase partition spatial data over the different DB engines on a geographic base. Consequently, range-queries with small area are likely served by a subset of DBs in case of OpenGeoBase, while are served by all the DBs in case of MongoDB, resulting in a lower latency. The increase in the range-query area increases the number of involved OGB engines, and OpenGeoBase latency gets lower than MongoDB one. The OGB inefficiency related to not (always) using all DB engines can be merely avoided randomly assigning tiles to different database engines, but in doing so the administrator would lose the control of where spatial data are stored,  which is a unique feature of OGB system. Anyway, the analysis of different tile assignment schemes are left for further studies.

\subsection{Application tests}                      
We exploited OpenGeoBase as a back-end for an Intelligent Transport System (ITS) application, which is designed to discover information (bus stops, schedules, train time-table, etc.) available  for a given geographical area \cite{webapp}. Such information is expressed in the GTFS format \cite{GTFS}, usually used by ITS applications such as OpenTripPlanner.  
 
We downloaded from the web about 1000 public GTFS files made available by transport agencies. Each file contains the GPS coordinates of the stops of an agency transport service (bus, metro, train, etc.). For each GTFS file we created a GeoJSON Multipoint object, where each point has the coordinates of a stop. Each GeoJSON object has a "URL" properties whose value is the Internet URL of the GTFS file. The resulting distribution of not-void tiles is rather spread out, as show in fig \ref{f:gtfsword}, and we use Bloom Filter (BF) service. 

We have 4 database engines serving 4 different zones of the world, running in  different servers. Each database engine has an ICN cache of 130k ContentObjects. The front-end library (fig. \ref{f:arch}) is a Java code running in a Spring STS Server. The Application is a JavaScript code running in the end-user browser \cite{webapp}, which submits the range-query to the front-end library that executes constrained tessellation with $k=50$, bloom pre-filtering and tile-queries. Then, it collects, post-filters and sends back results. Front-end, databases and BF server are different devices located on the same LAN. The tests have been carried out by generating square intersect range-queries with different areas, randomly centered over Europe. Performance are measured on the front-end device, not including JavaScript processing. 

Fig. \ref{f:gtfsqt} reports the average range-query time versus area. The plot also reports the average delays that compose the range-query delay, namely: tessellation time, the time to interact with the BF Server and the duration of the batch of ICN tile-queries composing the range-query. Post-filtering processing has a negligible time. For extremely large range-queries, e.g. 1,000,000 km$^2$, the queries last about 230 ms. Fig. \ref{f:gtfstiles} reports the average number of tessellated tiles versus the range-query area. For areas greater than 35,000 km$^2$, the constraint $k=50$ can not be respected and the tessellation resorts to cover the area with a linearly increasing number of 100x100 tiles. Figure also reports the number of tile-queries actually carried out after the interaction with the BF server. The tessellation processing time increases with the range-query area but for large areas it tends to decrease since the constraint $k$ can not be satisfied and in these cases tessellation processing is very simple.

\section{Conclusions}
We exploited ICN to realize distributed spatial databases, which can horizontally scale by deploying new servers. We showed in a real use case the benefits of ICN's routing-by-name, in-network caching and data-centric security. Clearly, convincing users of current products (such as PostGIS, MongoDB, etc.) to adopt ICN-based solutions requires more research work, performance analysis and dissemination effort, which in our opinion is worth undertaking.

%\section*{Acknowledgements}
%This work is supported in part by the European Commission in the context of the H2020 Bonvoyage project.

\bibliographystyle{IEEEtran}
\bibliography{paper}

\clearpage

\ifpaper
\else	
\section*{APPENDIX I - Indexing Design}
A key design choice of a spatial database is how to map the space in an internal data structure to improve the speed of spatial operations. The mapping strategy is known as indexing method. In general, an indexing method partitions the space in regions that can be further decomposed in sub-regions. The resulting region hierarchy forms a tree data structure. Most popular indexing methods are Grid, R-Tree, and their variants \cite{MicrosoftSQL}.
 
Grid methods decompose the space into a uniform grid. A region of the grid, called \emph{tile}, can be iteratively split into smaller tiles, realizing a multi-layer hierarchical grid structure. Tiles are the nodes of the tree indexing structure \cite{quadtree}. Range-queries can be carried out "tessellating" the requested window with grid tiles and then fetching tiles data. 

R-trees decompose the space in overlapping rectangles \cite{guttman1984r}, whose number, size and position depend on the stored spatial objects. Spatial objects are enclosed in rectangles, each rectangle is a node of the indexing tree and it is the minimum bounding rectangle that encloses its child rectangles; leaf rectangles contain only one spatial object. A range-query is carried out using a recursive algorithm: starting from the root node it go down in the tree.  

The general methodology we follow to realize an ICN distributed spatial database consists in assigning unique and routable name prefixes to the regions identified by the indexing process. Fig. \ref{f:grid} reports an example of such regions (R) in case of Grid and R-tree indexing schemes and point objects. Each database engine stores data of a subset of regions and exposes the region name-prefixes to the ICN routing plane. Such a solution can be applied both in case of Grid and R-tree indexing. However, OpenGeoBase uses a Grid index, as we believe that it is better suited to work with ICN. Indeed: 

\begin{itemize}
	\item regions of a Grid index do not change, i.e., do not depend on the stored objects. Thus, we can assign unique name-prefixes to them at the system roll-out, and ICN routing updates will be needed only in case of administrative reconfigurations, e.g. moving region data from an engine to another one. If regions were dynamic, like in the case of R-Tree, we would risk the introduction of new names in the system and routing plane updates at each data insert or removal;
	\item nodes of a Grid indexing tree have a unique parent node, easing the use of hierarchical naming schemes. In case of R-tree, a node of the index may have multiple parents and parents may change over time, thus complicating the use of a hierarchical naming scheme;	
	\item query-handler and insert-handler need to know the shape of regions, to resolve which are the regions involved in range-query and data insert operations. Such knowledge is implicit for Grid methods, and resolution can be autonomously performed by end-devices or application servers. In R-tree methods, the region shape is not known a-priori and can change; thus, the resolution algorithm needs to be either informed of changes, or realized in recursive way \cite{du2007sd}, so increasing the number of message exchanges involved in the resolution process and then the query time.
\end{itemize}

\begin{figure}[t]
	\centering
	\subfigure[Grid]{  
		\includegraphics[scale=0.30]{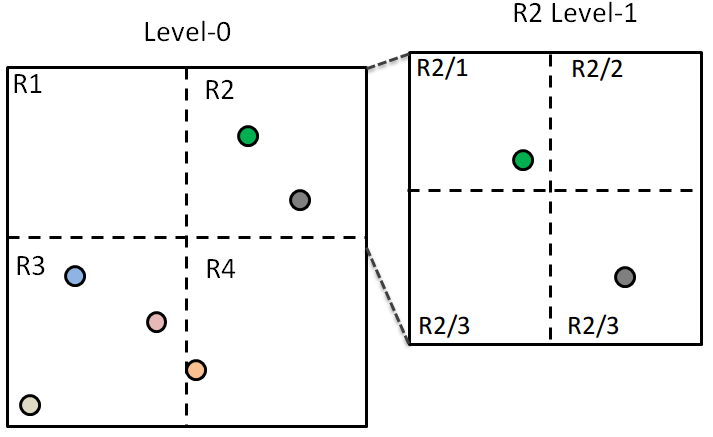}
		\label{f:grid}
	}
	\subfigure[R-tree]{  
		\includegraphics[scale=0.30]{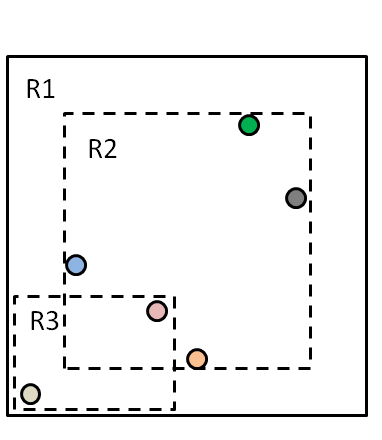}
		\label{f:rtree}
	}
	\caption{Regions of Grid and R-tree indexing schemes}
	%\label{f:indexing}
	
\end{figure}

\section*{APPENDIX II - Database Engine Internal Design}

A database engine processes tile-queries expressed by means of Interests of OGB-Tile contents, and stores geo-referenced data in the form of OGB-Data ContentObjects. To construct the engine, we extend the NDN Repo-ng \cite{ndn}. Repo-ng is formed by: a front-end module that processes Interest messages and TCP bulk-inserts; a back-end database management system (DBMS), in which it is possible to SELECT and INSERT ContentObjects through SQL transactions. The DBMS of NDN Repo-ng is SQLite and contains a single table (CO-Table), whose rows are the stored ContentObjects. 

We extend Repo-ng in order to handle and speed up the processing of tile-queries. We add three tables to the DBMS, named Tile-Tables, one for each OpenGeoBase spatial grid level. The rows of the level-$n$ Tile-Table contain the names of OGB-Data ContentObjects belonging to any tile of level-$n$. When a tile-query for a tile $x$ of level-$n$ is received, the front-end performs an SQL SELECT on the rows of level-$n$ tile-table, searching for names whose prefix is equal to the prefix of tile $x$. By having different tables per level, we reduce the search space and the tile-query processing time. Clearly, other possible table designs are possible, but here we wish to put forward the problem, rather than finding the best solution. The SQL SELECT returns to the Repo-ng front-end the names of ContentObjects, which are retrieved from the CO-Table. The Repo-ng front-end collects and sends them back to the querying devices, within one or more OGB-Tile ContentObjects, by using plain NDN segmentation. 
We also modify the Repo-ng TCP-bulk-insert procedure, so as to create related rows of tile-tables.

\begin{figure*}[t]
	\centering
	\subfigure[Range-query (pink box)]{ 
		\includegraphics[scale=0.26]{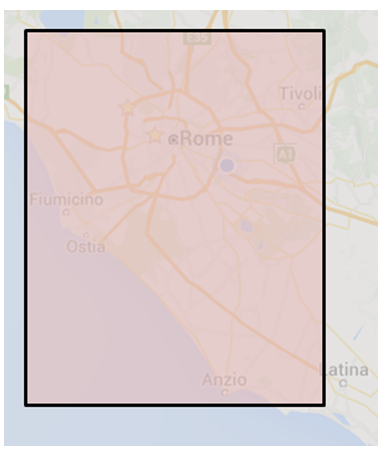}
		\label{f:range-query}
		
	}
	\subfigure[Min stretch tessellation]{ 
		\includegraphics[scale=0.26]{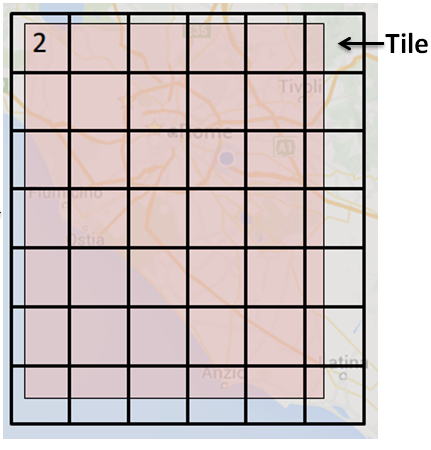}
		\label{f:min-streatch}
		
	}
	\subfigure[Min stretch-and-tiles tessellation]{ 
		\includegraphics[scale=0.26]{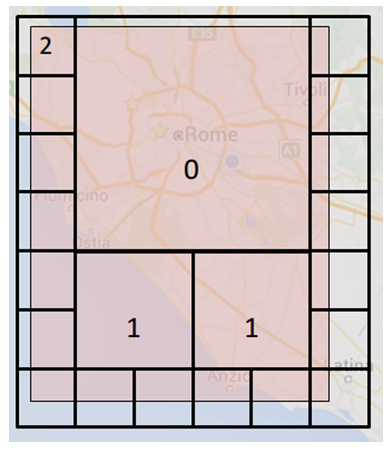}
		\label{f:min-streatch-and-tiles}
		
	}
	\subfigure[Constrained tessellation ($k=19$)]{ 
		\includegraphics[scale=0.26]{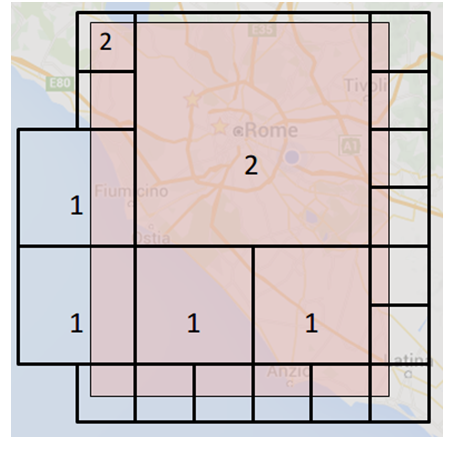}
		\label{f:constrained}
		
	}
	\caption{Range-query and tessellation approaches in case of grid indexing with three levels 0,1,2 and level-ratio 4}
	\label{f:indexing}
	
\end{figure*}

\section*{APPENDIX III - Tessellation Algorithm}
A range-query of an area $A$ is resolved with a set of tile-queries for tiles identified by the tessellation procedure. The set $T$ of tiles forming the tessellation covers an area $B$, which contains the area $A$, but also some additional border space, due the fact that the area $A$ may be not aligned with the grid (e.g., see fig. \ref{f:min-streatch-and-tiles}). We define \emph{tessellation stretch} $TeS$ the ratio between the area $B$ and the area $A$, namely $TeS = area(B)/area(A) \geq 1$.

A \emph{minimum stretch} tessellation is formed by the minimum set of the smallest tiles that intersects the area $A$, e.g. level-2 tiles as shown in fig. \ref{f:min-streatch}. However, this procedure may return many tiles, even for small values of $A$, which should be queried through ICN means, dramatically increasing the overall range-query time. For instance, using 1 km x 1 km tiles, a range query of 50 km x 50 km is tessellated at least with 2500 tiles, i.e. 2500 tile-queries are necessary to satisfy the range-query. Definitively too much.

To achieve a first reduction of the number of tiles, we can simply exploit the hierarchy of the grid, by removing each set of tiles of level-$n+1$ that completely fills the parent tile of level-$n$, and insert such a tile. This is for instance the case of the tessellation reported in fig. \ref{f:min-streatch-and-tiles}, in which a level-1 tile takes the place of 4 child level-2 tiles and, recursively, a level-0 tile replaces 4 child level-1 tiles.

\begin{figure}[t]
\centering
\includegraphics[scale=0.54]{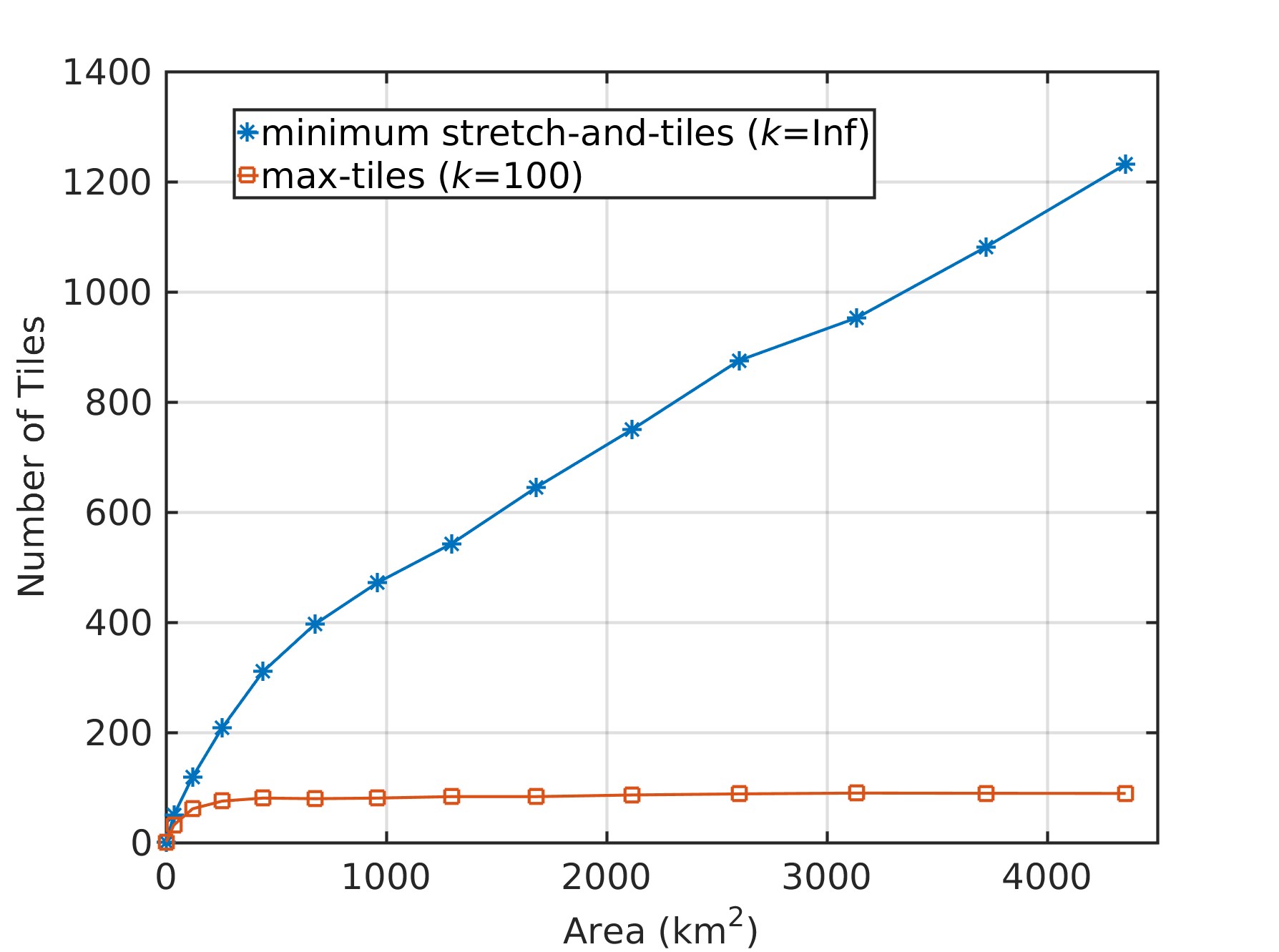}
\caption{Number of tiles vs. range-query area for minimal stretch-and-tiles tessellation for the OGB grid}
\label{f:tesseling-perf1}
\vspace{-8pt}
\end{figure}

However, even this \emph{minimum stretch-and-tiles} tessellation may return a high number of tiles, in case of large range-query, e.g. numbers of tiles greater than 500. Indeed, fig. \ref{f:tesseling-perf1} shows the number of tiles for randomly-centered range-query of square shape, whose area is the x-axis of the plot, by using the OpenGeoBase spatial grid. Such high values are usually due to the numerous small tiles used to fill the borders of the range-query area and/or to the misalignment of the range-query area with respect to the grid hierarchy. 

For this reason, we developed a \emph{constrained} tessellation algorithm, in which it is possible to fix the maximum number $k$ of returned tiles. For instance, fig. \ref{f:constrained} shows the result of a constrained tessellation for $k=19$. The drawback is that the lower the maximum number of allowed tiles, the higher the tessellation stretch. 

An optimal constrained tessellation algorithm should find the set of $k$ tiles providing the minimum stretch. This problem can be made equivalent to a minimal-weight size-constrained set cover problem. The set $S$ to be covered is formed by the set of the smallest tiles intersecting the area $A$; the universe $U$ is formed by these tiles and all their parent tiles up to level-0. The cost of a tile of $U$ is its \emph{tile-stretch}, i.e. the ratio between the tile area and the area of its intersection with the range-query area. 

In 2015 the authors of \cite{golab2015size} demonstrated that this problem is NP-hard. Thus, we devised the simple, but effective, greedy algorithm \ref{a:ct}, reported below, which follows a top-down approach, by first inserting the necessary bigger tiles and then, iteratively, the necessary smaller ones. A level-$i$ tile is necessary when completing the tessellation with all remaining level-$(i+1)$ tiles would not be able to respect the constraint $k$. Thus a level-$i$ "aggregation" is surely necessary. For very large range-query, it is not possible to respect the constraint, even by using all biggest tiles of level-0. In this case, the algorithm returns the set of covering level-0 tiles, even if it exceeds the constraint.

\begin{algorithm} [ht]
\caption{Constrained Tessellation}
\label{a:ct}
%{\fontsize{8}{10}
\begin{algorithmic}
\Statex $k$ = max number of tiles
\Statex $n$ = number of levels of the spatial grid
\Statex $S$ = whole indexing tree
\Statex $leaf(S)$ = leafs of S, i.e. tiles of the tessellation
\Statex $MST(S)$ = minimum stretch-and-tiles reduction of $S$
\Statex
%\Statex\textit{Tesseling}\

\Statex\textit{constraint violation exception}
\If {Using all level-0 tiles the number of tiles $> k$}

\Return the set of covering level-0 tiles

\EndIf
\Statex
\Statex $S=MST(S)$

\While{number of $leaf(S) > k$}

$i=0$

\While{$i \leq n$}
\If {(A new level-$i$ tile is necessary)}
\State Insert a new level-$i$ tile with min tile-stretch
\State Remove all its children
\State break
\EndIf

$i=i+1$

\EndWhile
\EndWhile\\		
\Return leaf(S) 
\end{algorithmic}
%}
\end{algorithm}

This algorithm is not optimal; however, we made some trials using a brute-force search of the optimal solution, and we found extremely rare cases in which our solution is not the minimum stretch one. In any case, finding the optimal constrained tessellation is not the aim of this paper, as we just wanted to design a working solution.

\begin{figure}[t]
\centering
\includegraphics[scale=0.54]{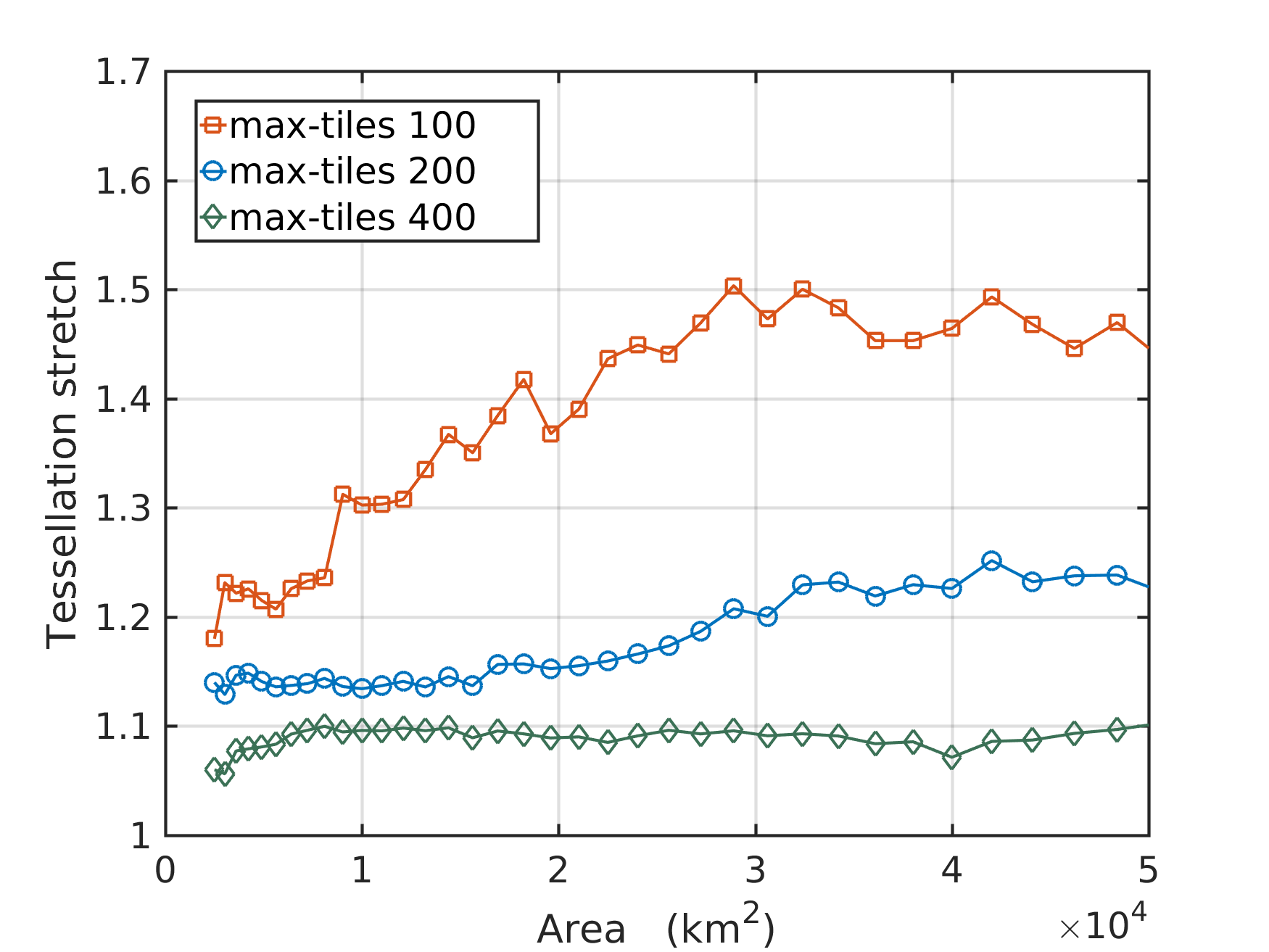}
\caption{Stretch of constrained tessellation vs. range-query area for the OGB spatial grid}
\label{f:tessellating-strech}
\vspace{-6pt}
\end{figure}

Fig. \ref{f:tesseling-perf1} shows that the constrained tessellation algorithm is actually able to reduce the number of queries up to the constraint $k=100$. Fig. \ref{f:tessellating-strech} reports the expected drawback: the increase of the tessellation stretch when the maximum number of tiles $k$ decreases. 
%Roughly, using 100 tiles we achieve have a stretch about of 50\% for very large query-areas, such as 50,000 km$^2$. Using 200 and 400 tiles this value fall down to 22\% and 10\%, respectively \footnote{We observe that also in case of the unconstrained minimal stretch-and-tiles solution there is a little stretch due to the fact that we can not have a resolution lower than the smaller grid level; however stretch value are extremely close to one}.     
With stretch greater than 1, a range-query returns geo-referenced information related to an area that is greater than the requested one, with a consequent increase of the network traffic. 
%which can be filtered out at the application end. 
Consequently, on the one hand, reducing the number of tiles accelerates the range-query processing time, since less ICN tile-queries are required; on the other hand, we are increasing the transmission time, since transferred data may increase. Clearly a trade-off is needed, which depends on the density and volume of geo-referenced information. Such values are application-dependent.

\fi

\end{document}